\documentclass[twoside]{article}
\usepackage{fleqn,espcrc2,epsf}

\usepackage{graphicx}
\usepackage[figuresright]{rotating}

\newcommand{\AmS}{{\protect\the\textfont2
  A\kern-.1667em\lower.5ex\hbox{M}\kern-.125emS}}

\title{Instanton Effects in Hadron Spectroscopy Revisited}

\author{Tam\'as G. Kov\'acs\thanks{Suppported by FOM, Dutch Research Fund.}
            \address{Instituut-Lorentz for 
                  Theoretical Physics, University of Leiden,\\ 
             P.O. Box 9506, NL-2300 RA Leiden, The Netherlands}}
       
\begin{document}

\begin{abstract}
We use an optimised clover action to study spectroscopy
on an instanton ensemble reconstructed from smoothed
Monte Carlo configurations. Due to the better chirality of the clover action, 
the artificial configurations show a marked difference from the free field
behaviour obtained with the Wilson action. They however still fail to 
reproduce the physics observed on the smoothed configurations.
The presence of freely propagating quark modes is found to
be responsible for this.
\vspace{1pc}
\end{abstract}

\maketitle

According to the instanton liquid model most of the
low-energy properties of QCD can be explained by 
instantons \cite{Shuryak}. 
This however has never been fully tested starting from first
principles, in particular on the lattice. In a previous study
hadron spectroscopy was performed on instanton ensembles
reconstructed from smoothed Monte Carlo configurations
\cite{Ispec}. The instanton ensemble was found to exhibit 
chiral symmetry breaking but the lightest states in the 
pion and rho channel remained degenerate down to small 
quark masses; a feature typical of free field theory
(no gluons).

Now we would like to shed some more light on the origin of
this peculiar admixture of ``free field like'' and ``QCD like''
properties of the instanton ensemble. We shall compare the 
following two ensembles of gauge configurations:
\begin{itemize}
\item{Quenched Monte Carlo gauge configurations produced with
the fixed point action of \cite{Ispec} at a lattice spacing
of $a=0.144$fm and then cycled (smoothed with a renormalisation 
group based procedure) 9 times, up to the point where instantons
could be reliably identified.}
\item{Artificially constructed instanton configurations 
reproducing the instanton content (location and size but not
the relative orientation in internal space) 
of the above (for more details on the 
construction, see Ref.\ \cite{Ispec}).}
\end{itemize}
We shall refer to the first (smoothed) and the second (instanton)
ensemble as SM and IN respectively. The SM ensemble is so smooth
that 70\% of its action is carried by the instantons (assuming
no interaction between them), still it has been shown to contain
all the relevant long distance features of QCD.

An important feature of the instanton liquid model is the presence
of near zero quark eigenmodes that are approximately linear combinations
of instanton and antiinstanton zero modes. Their formation is
possible only if the zero modes are (at least nearly) degenerate.
Only then can small perturbations --- e.g. gauge field
fluctuations --- mix them. This is not the case with the Wilson
action. Since it breaks chiral symmetry explicitly, the 
would be fermion zero modes spread in an
interval of the real axis. 

\begin{figure}[htb]
\begin{minipage}{75mm}
{\ }\hfill\epsfysize=6.0cm\epsfbox{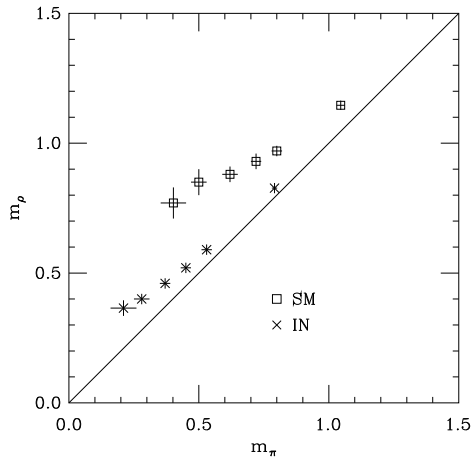}\hfill{\ }\\
\vspace{-1.2cm}\caption{\label{fig:pirho} $m_\rho$ versus $m_\pi$
obtained with the optimised clover action on the instanton 
ensemble (crosses) and on the smoothed ensemble (boxes).
The $m_\pi=m_\rho$ line is the free field result, given also
by the Wilson action on the SM ensemble.}
\end{minipage}
\vspace{-3mm}
\end{figure}
                            
To check how important the chirality of the fermion action is,
we repeated the spectroscopy with the clover action (c$_{sw}$=1.2)
optimised to produce ``zero modes'' in a narrow range around zero
\cite{Opt}. Fig.\ \ref{fig:pirho} shows the result 
for $m_\pi$ vs. $m_\rho$. This markedly differs from the
$m_\pi=m_\rho$ free field line (also shown) that was obtained
with the Wilson action. This clearly shows
the importance of the chiral symmetry of the fermion action.
On the other hand, the $\pi-\rho$ splitting on 
the IN ensemble still fails to reproduce that on the 
SM ensemble. In view of the good chirality
of the optimised clover action, it is hard to imagine that
this can be fully attributed to the remaining explicit
chirality breaking of the action.

\begin{figure}[htb]
\begin{minipage}{75mm}
{\ }\hfill\epsfysize=6.0cm\epsfbox{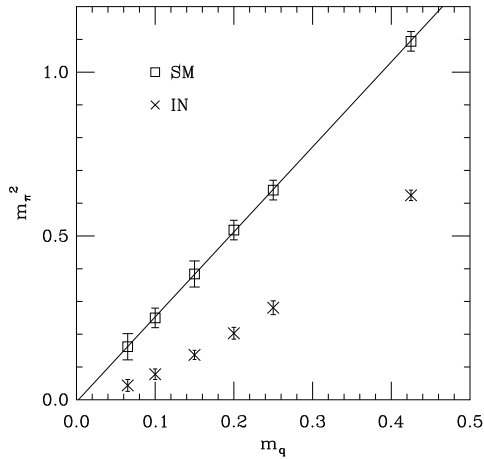}\hfill{\ }\\
\vspace{-1.2cm}\caption{\label{fig:pi} $m_\pi^2$ versus 
the bare quark mass, $m_q$ for the optimised clover action 
on the instanton ensemble (crosses) and on the smoothed ensemble (boxes).
The line is the best fit to the SM data.}
\end{minipage}
\vspace{-3mm}
\end{figure}

Fig.\ \ref{fig:pi} shows the pion mass sqared vs.\ the bare 
quark mass for both ensembles. On the SM ensemble $m_\pi^2 \propto
m_q$, as expected from PCAC.
This is clearly not the case on the IN ensemble, where a 
best fit to the form $m_\pi^\lambda \propto m_q$ gives $\lambda=1.5$
which is between $\lambda=2$ and the free field value, $\lambda=1$.

To obtain a more ``microscopic'' understanding of the nature of
quark propagation in the two ensembles, we also looked at some 
randomly selected close to zero quark eigenvalues and the
corresponding eigenmodes. 
                           
\begin{figure}[htb]
\begin{minipage}{75mm}
{\ }\hfill\epsfysize=6.0cm\epsfbox{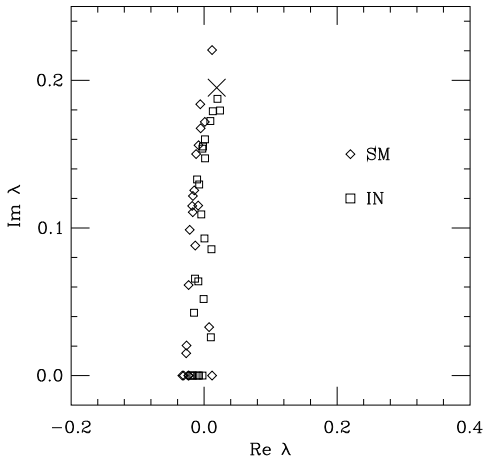}\hfill{\ }\\
\vspace{-1.2cm}\caption{\label{fig:eigen} The 
eigenvalues of the optimised clover Dirac operator
on the smoothed (diamonds) and the instanton (boxes) 
ensemble. The cross shows the lowest
antiperiodic free field mode.}
\end{minipage}
\vspace{-3mm}
\end{figure}
                            
On a flat gauge field configuration with periodic bondary 
conditions in all directions, there are $4N_c$ trivial 
(constant) zero modes, this is the number of c-number
degrees of freedom corresponding to one fermion flavour.
In the presence of antiperiodic boundary condition in the 
time direction, the lowest eigenmodes are shifted away 
from zero, to
\begin{equation}
\mu_\pm = 1 - \cos\frac{\pi}{N_t} \pm i sin\frac{\pi}{N_t} =
0.019 \pm 0.195 i,
\end{equation}
since in our case $N_t=16$. Both eigenvalues are $4N_c$-fold
degenerate. For purely technical reasons we
use antiperiodic boundary condition in the time direction.
This separates the lowest free field modes from the lowest 
topological modes and makes the computation faster.
For orientation, in Fig.\ \ref{fig:eigen} we plotted 
the eigenvalues that we used from both ensembles, together
with the location of the lowest antiperiodic free field mode.

The mixing of the free field modes into the eigenmodes 
of SM and IN can be characterised by $P_\pm(\lambda) =
\| P_\pm \psi_\lambda \|$, where $P_\pm$ is the projection of
the normalised eigenmode $\psi_\lambda$ onto the eigenspace 
corresponding to the eigenvalue $\mu_\pm$. This quantity is
not gauge invariant therefore we work in Lorentz gauge.
Any generic fermion mode on a non-flat 
but locally smooth configuration will have a nonzero 
projection on the $\mu_\pm$ eigenspaces. The question is 
whether this is just an accidental mixing or the given
ensemble really contains close to free field modes. To decide
this, a good quantity to look at is $P(\lambda)=
P_+(\lambda)/P_-(\lambda)$. If the mixing is accidental, we
expect $P(\lambda)$ to fluctuate around 1, independently
of the corresponding eigenvalue $\lambda$. On the other
hand, if there are free field like modes on a given 
configuration then $P(\lambda)$ will increase substantially
when $\lambda$ approaches the value $\mu_+$. 
 
\begin{figure}[htb]
\begin{minipage}{75mm}
{\ }\hfill\epsfysize=6.0cm\epsfbox{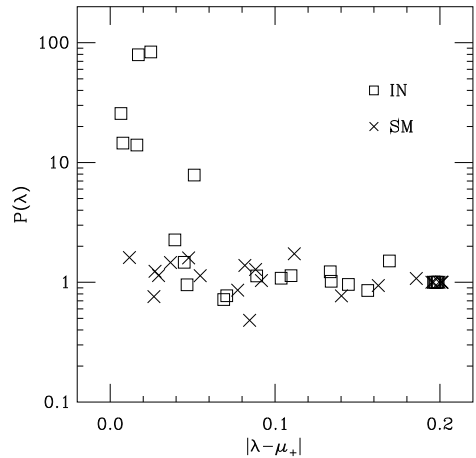}\hfill{\ }\\
\vspace{-1.3cm}\caption{\label{fig:proj} 
$P(\lambda)=P_+(\lambda)/P_-(\lambda)$ versus the distance
of the eigenvalues $(\lambda)$ from the free field eigenmode
$\mu_+$ for the instantons only (boxes) and the smoothed 
(crosses) ensemble.}
\end{minipage}
\vspace{-3mm}
\end{figure}

In Fig.\ \ref{fig:proj} we plot $P(\lambda)$ versus the
distance of the given eigenvalue from the free field mode
$\mu_+$. In the SM ensemble $P(\lambda)$ 
fluctuates around 1 everywhere, there is no trace of the
free field mode. On the other hand, in the IN ensemble,
the modes close to $\mu_+$ have a substantially larger 
projection on the $\mu_+$ eigenspace than on the $\mu_-$
subspace. In fact, these modes, close to $\mu_+$ have 
$\|P_+(\lambda)\| \approx 1$, so
they are essentially free field modes. A detailed study of
the chiral density $\psi_\lambda^\dagger \gamma_5 \psi_\lambda$
reveals that all the $P(\lambda)\approx 1$ modes look like
mixtures of instanton zero modes with the density 
concentrated in several lumps. This is to be contrasted with
the chiral density of the modes with $P(\lambda) \gg 1$
that spreads roughly homogenously over the whole lattice,
as expected of the lowest free field eigenmode.

In conclusion, we found that the instanton ensemble 
reconstructed from smoothed Monte Carlo configurations
shows a peculiar mixture of free field and interacting
QCD like behaviour. This is a consequence of the presence
of both types of quark eigenmodes in this ensemble.
The scattering of quarks by instantons is apparently not enough
to eliminate free propagation completely.
In contrast, we have not found any trace of the free field
modes on the smoothed configurations. This is presumably 
due to some long wavelength fluctuations also responsible for 
confinement which is absent from the IN ensemble. 

Finally we note that the density of modes around zero
goes as $\propto V^{1/4}$ in free field theory, whereas
the density of instanton modes in the instanton
ensemble is $\propto V$, the lattice
volume. It is thus concievable that in the absence of confinement,
there are always some free field modes which contribute
to quark propagation but in the $V \rightarrow 
\infty$ limit they might be overwhelmed by the
instanton modes. It would be interesting --- although probably
not very cheap --- to test this directly on the lattice.
The other possibility is that instantons alone might
not be enough to reproduce low energy QCD, confinement
is also needed. In this case it is confinement
that eliminates free quark propagation and
makes the hopping of quarks from instanton
to instanton the dominant long distance mode of propagation.

\end{document}